\documentclass[reqno,11pt,a4]{amsart}
\usepackage{amsmath,amssymb,tabularx,setspace,color}
\RequirePackage[OT1]{fontenc}
\usepackage{multirow}
\newtheorem{theorem}{Theorem}[section]
\newtheorem{lemma}{Lemma}[section]
\numberwithin{equation}{section}
\usepackage{graphicx,graphics,enumerate}
\usepackage{epstopdf}
\usepackage{graphicx}
\usepackage{caption}
\usepackage{subcaption}

\makeatletter
\renewcommand\section{\@startsection {section}{1}{\z@}%
                                   {-3.5ex \@plus -1ex \@minus -.2ex}%
                                   {2.3ex \@plus.2ex}%
                                   {\normalfont\large\bfseries}}
\makeatother
\newtheorem{Example}{Example}
\newtheorem{Theorem}{Theorem}
\newtheorem{Remark}{Remark}[section]

\newtheorem{Definition}{Definition}[section]

\begin{document}
\doublespace
\title[]{Semiparametric  transformation Model with  measurement error in Covariates: An Instrumental variable approach}
\author[]%
{S\lowercase{udheesh} K. K.$^{\lowercase{a}\dag}$,  D\lowercase{eemat} C. M\lowercase{athew}$^{\lowercase{b}}$, L\lowercase{itty} M\lowercase{athew}$^{\lowercase{a}}$ \lowercase{and} M\lowercase{in} X\lowercase{ie}$^{\lowercase{c}}$\\
 $^{\lowercase{a}}$I\lowercase{ndian} S\lowercase{tatistical} I\lowercase{nstitute},
  C\lowercase{hennai}, I\lowercase{ndia,}\\
  $^{\lowercase{b}}$S\lowercase{t.} T\lowercase{homas} C\lowercase{ollege}, P\lowercase{ala}, I\lowercase{ndia}, \\
  $^{\lowercase{c}}$C\lowercase{ity} U\lowercase{niversity of} H\lowercase{ong} K\lowercase{ong},  H\lowercase{ong} K\lowercase{ong}.
}
\thanks{ {$^{\dag}$}{Corresponding E-mail: \tt skkattu@isichennai.res.in}} %The part of this work is done while Sudheesh K K was visiting City University of Hong Kong.}
\maketitle
\begin{abstract}
The linear transformation  model provides a general framework for analyzing  censored survival data with covariates. The proportional hazards and proportional odds  models are special cases of the linear transformation  model. In biomedical studies, covariates with measurement error may occur in survival data. In this work, we propose a  method to  obtain estimators of the regression coefficients in the linear transformation  model when the covariates are subject to measurement error. In the proposed method, we assume that instrumental variables are available. We develop  counting process based estimating equations for finding the estimators of regression coefficients.
 We prove the large sample properties of the estimators using  the martingale representation of the regression estimators.  The finite sample performances of the estimators are evaluated through an extensive Monte Carlo simulation study. Finally, we illustrate the proposed method using an AIDS clinical trial (ACTG 175) data.
\end{abstract}

\section{Introduction}
Right censored survival data often arise in biomedical studies, clinical trials   and many other fields.
 The Cox  proportional hazards model  (Cox, 1972, 1975) has been fully explored
and widely used in the analysis of survival data which incorporates covariates.  The proportional odds model is an another
 approach used to find the effect of covariates on lifetime. (Bennett, 1983).
 %Pettitt, 1982).
 % , Bennett (1983) and Dabrowska and Doksum (1988)).
 Linear transformation model provides a general framework for analyzing the
 censored survival data and it relates  the transformed survival data to the covariates, which provides many useful
 models  including  Cox proportional hazards model and
 proportional odds model (Doksum, 1987). For the analysis of the linear transformation model, we refer interested readers to Cheng et al. (1995), Fine et al. (1998),  Chen et al. (2002),  Zeng and Lin (2006), Zeng et al. (2016),  Deresa and Van Keilegom (2021) and Zhou et al. (2022). Among these,  Deresa and Van Keilegom (2021) developed methods for analysing semiparametric linear  transformation model when the  lifetime is conditionally dependent of the censoring time, given the covariates. Zhou et al. (2022) developed a R package `TransModel' for linear transformation model with censored data, which helps practitioners to use the model easily and effectively.

In practice, accurate values of the covariates may not be available, instead one observes surrogate values of true covariates. For example, in biomedical studies, when protein intake for last 2 years becomes one of the variable of interest,  it can not be measured accurately. Similarly, in clinical studies on HIV, one important covariate is CD4 count, a measure of health of the immune system, will be subjected to  measurement error due to both imperfect instruments and biological fluctuation. In these situations, we need to analyze the data with surrogates.  Standard data analysis tools ignoring the fact of  measurement  error in covariates  may result bias and inconsistency in the
parameter estimation.  The estimation of underlying parameters in transformation model  when the independent variables are measured with error  have been rigorously carried out in the statistical literature (Hu and Lin (2004), Zucker (2005), Sinha and Ma (2014), Huang and Wang (2018), Li et al. (2019), Chen (2020)  and Chen and Grace (2020)).

 The estimators of regression coefficients obtained by plug in the surrogates instead
 of the true covariates, will be severely biased. Several methods have been
developed in the last two decades to remove the bias by adjusting the likelihood
function using a corrected term, which takes into account of the measurement error.
The corrected score method generally assumes the availability of a validation sample
%(Wang et al. (1997), Zhou and Wang (2001))
for true and mis-measured covariates. Wang and Wang (2015) proposed a method to find the  estimators of the regression coefficients for the linear transformation model with covariate measurement error in the presence of  validation sample. One needs to make assumptions about the measurement error process (Hu et al.,  1998) when
a validation sample is not available. Hu et al. (1998) developed a likelihood based
method for analyzing proportional hazards model based on the assumption of  distribution
of the true covariates.  This assumption enables us to find the measurement error
variability. Cheng and Wang (2001) analyzed the linear transformation model under the
assumption that the measurement error variance is known. Bertrand et al. (2019) proposed a flexible likelihood based  approach to estimate the   additive measurement error variance. They illustrated the use of the
estimator of the measurement error variance using Cox proportional hazards model.   Apart from these, the analysis
of the linear transformation model with covariates measured with error  is possible, if we have repeated measurements for mis-measured covariates, which allow the estimation of measurement error variability (Song and Wang (2014) and Sinha and Ma (2014)).

   Sinha and Ma (2014) developed a semi-parametric approach to analyze the linear
transformation model with measurement error, when repeated measurements of a
surrogate of the unobserved true predictor are available.    Mandal et al. (2019)  proposed a general  approach  for the  analysis of the semiparametric  linear transformation model with  interval censored data when the covariates are measured with error.  Semiparametric regression for measurement error model with heteroscedastic error has been studied by Li et al. (2019). They proposed a method based on efficient score functions to deal with the heteroscedasticity of the measurement error.  Chen (2020) proposed  corrected estimating equation using  the
simulation-extrapolation (SIMEX) method for estimating the regression parameters in the linear transformation model with length biased data and covariate measurement error. Chen and Yi  (2021) developed augmented nonparametric maximum likelihood estimation method for the proportional hazards model  for handling left truncated and right censored survival data with measurement error in the covariates.

The existing literature on
measurement error models with linear transformation generally require the knowledge of
either repeated measurements of error prone covariates or validation data set, which  enable
us to estimate the measurement error variance. The availability of repeated measurements or
validation data set are not always realistic. Moreover, in the absence of
information about the measurement error variance, estimators of the regression parameters
can be obtained if instrumental variables are available. For example, consider the CD4 measurements of patients taken during HIV studies. The CD4 counts should be measured at different time points to check the variation from the baseline inorder to determine the study outcome. In an AIDS clinical Trial group (ACTG 175) study conducted by Hammer et al. (1996), one of the study end point was the time at which greater than 50\% decline in CD4 cell count. The objective of the study was to assess the effect of treatments on survival time adjusted for baseline CD4 counts. The CD4 measurement within one week before randomization was taken as the baseline CD4 measurement. Since the observed CD4 counts might be affected by measurement error, use of an instrumental variable can reduce the bias in the estimates of regression coefficients. Here, CD4 measurement between one to three weeks prior to randomization can be used as an instrumental variable.

The instrumental variable approach has got special  attention in the measurement error analysis
literature, due to its methodological
flexibility and practical applicability. For more details refer to Schennach (2007), Wang and Hsiao (2011), Wang and Song (2016), Guan et al. (2019) and  Wei et al. (2021) and the references therein. An instrumental variable is an additional measurement of covariate that satisfy
three requirements: i) Instrumental variable is non-differential, ii) It is correlated with the
model covariate and iii) It is independent of the measurement error. The variable that  is highly correlated with unobserved covariate but independent of measurement error can be used as an
instrumental variable.  Note that a replicate observation is an instrumental variable but an instrumental variable is not
necessarily a replicate (Carroll et al. 2006). It is possible to use an instrumental variable to
estimate the measurement error variance. The main advantage of the instrumental variable
method is that it does not require any assumption regarding the distribution of the mis-measured
covariate and of the measurement error. Schennach (2007) and Wang and Hsiao (2011)
showed that measurement error linear regression models are identifiable, when instrumental
variables are available for the study.

Under the proportional hazards model assumption, Song and Wang (2014) developed a simple nonparametric correction approach for estimating the regression coefficients  using a subset of the sample where instrumental variables are observed. They also proposed a generalized method of moments nonparametric correction estimator to improve the efficiency over their simple correction approach.   Using instrumental variable, Huang and Wang (2018) proposed a novel method for Cox proportional hazards model under the dependence between the measurement error and true covariates. Even though the instrumental variable approach is explored  in the  Cox proportional hazards model, measurement error analysis  using instrumental variable is not available for  linear transformation model. Given the importance of the linear transformation  model with measurement
error in the covariates, it is highly desirable to develop methods for finding the estimators of the regression
parameters in the presence of instrumental variables.

  The rest of the manuscript is organized as follows. In Section 2, We discuss how the instrumental variable approach can be used to estimate the regression parameters in the  linear transformation model. Asymptotic properties of the estimators are studied in Section 3. A Monte Carlo  simulation study is conducted to assess the finite sample performance of the  estimators and the result is reported in Section 4. In Section 5, We illustrate the proposed  method using ACTG 175 data. Concluding remarks along with some open problems are given in Section 6.

\section{Estimation of regression parameters }
We describe  the estimation of regression parameters in the linear transformation model with measurement error in covariates, in the presence of instrumental variables. Let $T$ be the survival time and $X$ be the  $p$-dimensional covariate vector.
The linear transformation model is specified by
\begin{equation}\label{eq2.1}
l(T)=-X\beta+e,
\end{equation}
where $e$ is a random error with a known distribution function $G(.)$ and is independent of $X$.  Here $l(.)$ is a completely unspecified strictly increasing function and  $\beta$ is an unknown $p$-dimensional vector of  regression coefficients.
The linear transformation model includes many useful models as special cases. For example,
if $e$  follows  standard extreme value distribution,  model (\ref{eq2.1}) becomes
the Cox proportional hazards model (Cox, 1972) and it  reduces to the proportional odds model, when $e$  has standard logistic distribution.

In practice, $X$ may not be observed accurately  and one needs to develop regression model to incorporates the error in $X$. Let $T_i$ be the failure time of the $i$-th individual, $i=1,2,...,n$ and let $X_i=(X_{i1},X_{i2},...,X_{ip})$ be the vector of covariates subject to measurement error.  The linear transformation model with error in covariates is specified by
\begin{equation}\label{LTM}
l(T_i)=-X_i\beta+e_i.
\end{equation}  Here $X_i$ is unobservable, instead one observes $$Z_i=X_i+v_i,$$ where $v_i$'s are independent and identically distributed  $p$-dimensional measurement error vectors.  We assume  each component of $v_i$ is symmetrically distributed with  mean zero and variance $\sigma_{v}^{2}$ and  $v_i$ is independent of $(T_i,X_i)$ for all $i$.  Denote $l(T)=(l(T_1),\ldots,l(T_n))^T$, $X=(X_1,\dots,X_n)^T$.  In matrix form,  model (\ref{LTM}) can be written as
 \begin{equation}\label{LTMM}
l(T)=-X\beta+e,
\end{equation}
\begin{equation*}%\label{eq3.2}
Z=X+v,
\end{equation*}where $Z$ and $v$ are  $n\times p$ matrices.   Transformation models without measurement error in covariates have been studied by Cheng et al. (1995), Chen et al. (2002) and Zeng and Lin (2006) among others.

Next, we discuss the estimation  of the regression parameter vector in randomly censored linear transformation model with measurement error in covariates. The estimation consists of two steps. In the first stage we estimate the measurement error variables using standard least square method in the presence of instrumental variables. In the second stage,  we obtain a partial likelihood by substituting the estimators of covariates and then we find the estimators of the regression parameters by maximizing the partial likelihood. Assume that vectors of instrumental variables $W_i \in R^q; i = 1,2,\ldots n; q \geq p,$ are available for the study and $W$ denotes the $n\times q$ matrix whose $i$-{th} row
is $W_i^T$.  Assume that $W$ is related to $X$ (Wang and Hsiao, 2011)
through
\begin{equation}\label{eq3.3}
 X = WQ + \varepsilon,
\end{equation}
where $Q$ is a $q \times p$ matrix of unknown parameters with rank $p$ and $\varepsilon$ is a $n \times p$ matrix
of measurement errors, independent of $W$ and $v$  with $E(\varepsilon) = 0$ and variance $\sigma_{\varepsilon}^{2}$.
In summary, the model of interest is specified by the following equations;
\begin{equation}\label{3.4}
l(T)=-X\beta+e, \quad Z=X+v, \quad   X = WQ + \varepsilon,
\end{equation}
where the problem of interest is the estimation of $\beta$ based on the observed  data. The above equations yield
\begin{equation}\label{eq3.5}
l(T)=-WQ\beta + \beta \varepsilon + e, \quad Z = WQ+\varepsilon+v.
\end{equation}

Next, we estimate the parameter  $Q$ in (\ref{eq3.3}).  Assume that the rank of $W$ is $q$ so that $W^TW$ is invertible. Note that under the model
assumptions, $E(Z|W) = WQ$. Hence the least square estimator of $Q$ is
\begin{equation}\label{qhat}
\widehat{Q}:= (W^TW)^{-1}W^TZ.
\end{equation}
Clearly, $\widehat{Q}$ is an unbiased estimator of $Q$. It is also consistent under the assumed model
conditions and under the assumption $W^TW/n$ converges in probability to a positive definite
matrix.    Next, we  develop a  martingale based estimation equation to find the estimators of regression  parameters.

Let $\tilde T_i=\min(T_i,C_i)$ and $\delta_i=I(T_i\leq C_i)$,  $i=1,2,...,n$, where $C_i$  is the censoring time.  We assume that conditional on the complete covariates $X_i$,  $T_i$ and $C_i$ are independent. Let  $N_i(t)=I(T\leq t,\delta_i=1)$  be the number of event observed for $i$-th subject by time $t$.  We denote $Y_i(t)$ as  the indicator that the $i$-th subject is under observation at time $t$. i.e. $Y_i(t)=I(T_i\geq t)$. Also let $\lambda(t)$  and $\Lambda(t)$ be the hazard rate and cumulative hazard rate of $e$, respectively. Then the cumulative intensity function corresponds to the model (\ref{LTM}) is given by
\begin{equation}\label{eq3.7}
\Lambda_{e_i}(t)=\int_0^t Y_i(u)d\Lambda(X_i\beta+l(u)),
\end{equation}
where $\Lambda_{e_i}(.)$ is the conditional cumulative hazard rate of $e_i$ given $X_i$.

Define
\begin{equation}\label{eq3.8}
M_{i}(t)=N_{i}(t)- \int_{0}^{t} Y_i(u)d\Lambda((WQ)_i\beta_0+l_{0}(u)),
\end{equation}
where $(\beta_0, l_0)$ are the true values of $(\beta, l)$ and $(WQ)_i$ denotes the $i$-th row of the matrix $WQ$.    Using the counting process and its associated martingale theory (Anderson and Gill, 1984), one can show that $M_{i}(t)$ is a mean zero martingale process with respect to appropriate filtration. Hence, the estimators of  $\beta$ and $l$ are  obtained  by solving  the following estimating equations.
\begin{equation}\label{eq3.9}
  U_{1}(\beta,l,\widehat{Q}) =\sum_{i=1}^{n}\int_0^\infty (W\widehat{Q})_i \big[dN_{i}(u)- Y_i(u)d\Lambda((W\widehat{Q})_i\beta+l(u))\big]=0,
\end{equation}
\begin{equation}\label{LTMO}
U_{2}(\beta,l,\widehat{Q}) =\sum_{i=1}^{n}\big[dN_{i}(u)- Y_i(u)d\Lambda((W\widehat{Q})_i\beta+l(u))\big]=0,\,u\geq 0.
\end{equation}Here, we use (\ref{qhat}) to find $\widehat{Q}$, the least square estimator of $Q$.
The partial likelihood  equations (\ref{eq3.9}) and (\ref{LTMO}) reduce to the estimating equations given by Chen et al. (2002), if the covariates are measured accurately. The partial likelihood equation reduces to the partial likelihood for Cox-proportional model when $\Lambda(t)=\exp(t)$.

 Given $\beta$, the left-hand side of the equation (\ref{LTMO}) is monotone in $l$ and  has a unique solution, denoted by $\hat{l}$. After  substituting $\hat{l}$ in (\ref{eq3.9}), one can solve the same equation to obtain the estimator $\widehat{\beta}$ of $\beta$.

 % Under certain regularity conditions, it can be shown that asymptotic distribution  of $\widehat{\beta}$ has normal distribution which is discussed in next section.
\section{Asymptotic Properties}

In this section, we obtain the asymptotic normality of $\widehat{\beta}$. We consider a martingale representation of  $\widehat{\beta}$
to find the asymptotic distribution.   We prove the asymptotic properties of  $\widehat{\beta}$ under some design and regularity conditions on instrumental variables.

Let $\tau= \inf\{t : P(T >t)=0\}.$ Assume  $\psi(t)= \frac{\partial}{\partial t} log \lambda(t)=\frac{1}{\lambda(t)}\frac{\partial \lambda(t)}{\partial t}$ is continuous. Also assume  $\lim\limits_{s\rightarrow\infty}\lambda(s)=0=\lim\limits_{s\rightarrow\infty}\psi(s)$ and  $l$ has continuous positive derivatives.  Assume the following conditions;
\begin{enumerate}[D1.]
 \item $Z$ and $W$ are bounded in probability.
\item  The parameter $\beta$ lies in a compact set of $\mathbb{R}^p$
 \item $W^TW/n$ converges a.s. to a positive definite matrix.
\item  The fourth moment of instrumental variables is finite, i.e. $E(||W||^4) < \infty$; where for
any matrix $A,$ $||A||$ denotes the Euclidean norm of the matrix $A$.
\item The counting process martingales defined in section 3 satisfies the regularity conditions as in Fleming and Harrington (1991), ensuring the martingale central limit theorem.
\item $P(Y(\tau)=1)>0$, $P(T>\tau)>0$ and $P(C=\tau)>0$.
\end{enumerate}

We further introduce some notations. For any $s<t \in (0, \tau]$, define
\begin{eqnarray*}
\noindent B(t,s)&=&\exp\left(\int_{s}^{t}\frac{E[\frac{\partial\lambda}{\partial t}( (WQ)_i\beta_0+l_{0}(u))Y(u)]}{E[\lambda ((WQ)_i\beta_0+l_{0}(u))Y(u)]}dl_0(u)\right)
\end{eqnarray*}
 Let %$\tilde{Z}=(Z,WQ)$ and
 $$ C(t)=E[WQ\lambda ((WQ)_i\beta_0+l_{0}(T))Y(t)B(t,T)]$$ and $$
  C_{d}(t)=E[\lambda ((WQ)_i\beta_0+l_{0}(T))Y(t)].$$ Denote   $ \mu(t)=\frac{C(t)}{C_{d}(t)},\,\mu_{W}(t)$.
  $$\Sigma_{\beta}=\int_{0}^{\tau}E\{(WQ-\mu(t))^T(WQ){\lambda}'\{(WQ)\beta+l_0(t)\}Y(t)\}dl_0(t),$$ where prime denotes the differentiation with respect to $t$.
  $$\Sigma_{Q}=\int_{0}^{\tau}E\{(W-\mu_W(t))^T(WQ){\lambda}'\{(WQ)\beta+l_0(t)\}Y(t)\}dl_0(t).$$
  \begin{eqnarray*}	\Sigma_2&=&\int_{0}^{\tau}E\{(WQ-\mu(t))(WQ-\mu(t))^T\lambda\{(WQ)\beta_0+l_0(t)\}Y(t)\}dl_0(t)\\
		\end{eqnarray*}
$$\Sigma_{12}=\int_{0}^{\tau}E\big((WQ-\mu_{\widehat Q}(t))(W\widehat Q-\mu_{\widehat Q}(t))^T\lambda(WQ\beta_0+l_0(t))Y(t)\big)dl_0(t)$$

  We now establish some preliminary results useful for studying the asymptotic properties of the regression estimators and the proof of these results are given in Appendix.
  \begin{lemma} Suppose $a>0$ and $b$ are fixed finite
numbers and define $\lambda^* (l_0(t))=B(t,a)$ and $\Lambda^*(x)=\int_{b}^{x}\lambda^*(u)du$. Then we have
  \begin{eqnarray*}
\frac{\partial}{\partial Q}\widehat{l}(t)&\stackrel p\rightarrow&-
  \int_{0}^{t}\frac{\lambda_{0}^{*}(l_0(s))}
 {\lambda^{*}(l_0(\tilde{T}_i))B_2(s)}E\big[W\beta{\lambda}'(WQ\beta+l_0(s))Y(s)\big]dl_0(s). \\
     \frac{\partial}{\partial \beta}\widehat{l}(t)&\stackrel p\rightarrow&-\int_{0}^{t}\frac{\lambda_{0}^{*}(l_0(s))
     }
 {\lambda^{*}(l_0(\tilde{T}_i))B_2(s)}E\big[WQ{\lambda}'(WQ\beta_0+l_0(s))Y(s)\big]dl_0(s).
   \end{eqnarray*}
  \end{lemma}
\noindent We use Lemma 3.1 to establish the following results.
  \begin{lemma}\label{lemma2}
    \begin{eqnarray*}
 \frac{1}{n}\frac{\partial}{\partial \beta}U_{1}(\beta,\widehat l,Q)_{\beta=\beta_0}\stackrel{p}\rightarrow -\Sigma_{\beta_0},
\end{eqnarray*}
  \end{lemma}
 \begin{lemma}\label{lemma3}
    \begin{eqnarray*}
  \frac{1}{n}\frac{\partial}{\partial Q}U_{1}(\beta,\widehat l,Q)_{Q=Q_0}\stackrel{p}\rightarrow -\Sigma_{Q_0}.
  \end{eqnarray*}  \end{lemma}
  \begin{lemma}\label{lemma4}
  Under the condition $D3$ we have
  \begin{equation*}
    \|\widehat Q - Q\|=O_p(1/\sqrt n).
  \end{equation*}
  \end{lemma}
\noindent Lemmas 3.2 and 3.3 are used to establish the asymptotic distribution of $\widehat\beta$, while Lemma 3.4 is needed to establish the CLT for $\widehat Q$.

\begin{theorem}
	Under the regularity conditions $D1-D6$, $\widehat\beta$ is consistent for $\beta$.  As $n\rightarrow\infty$, the distribution of $\sqrt{n} (\widehat\beta-\beta_0)$ is multivariate normal with zero mean vector  and variance-covariance matrix $\Sigma_{\beta_0}^{-1}\Sigma(\Sigma_{\beta_0}^{-1})^T$, where $\Sigma=\Sigma_1-\Sigma_2+2\Sigma_{12}$ with $\Sigma_1=\Sigma_{Q_0}(W^TW)^{-1}\Sigma_{Q_0}^T\sigma_{\eta}^{2}$ and $\sigma_{\eta}^2=\sigma_{\varepsilon}^2+\sigma_{v}^2$.
\end{theorem}

%Note that $\widehat{\beta}$ and $\widehat l$ are the solutions of   (\ref{LTMO}) and (\ref{eq3.9}), respectively.

   A consistent estimator of the variance-covariance matrix $\Sigma_{\beta_0}^{-1}\Sigma(\Sigma_{\beta_0}^{-1})^T$  can be obtained  by replacing $\beta$, $l$ and $Q$ with the consistent estimators  $\widehat\beta_{0}$, $\widehat l_{0}$   and
	$\widehat Q$,  respectively.

%
%..................Add some more.

\section{Computational Algorithm and Simulations}
The estimators $\widehat\beta$ and $\widehat l$ are obtained as the values that solve the estimating equations  (\ref{eq3.9}) and (\ref{LTMO}), respectively. For computational simplicity, we express the equations (\ref{LTMO}) in an alternative form.   Let $t_1,t_2,\ldots,t_K$ be the observed failure times and (\ref{LTMO}) can be expressed  as
\begin{eqnarray}\label{eq5.1}
% \nonumber % Remove numbering (before each equation)
\sum_{i=1}^{n}Y_i(t_1)\Lambda((W\widehat{Q})_i\beta+l(t_1))&=&1
  \end{eqnarray}
\begin{eqnarray}\label{eq5.2}
% \nonumber % Remove numbering (before each equation)
\sum_{i=1}^{n}Y_i(t_k)\big(\Lambda((W\widehat{Q})_i\beta+l(t_k))-\Lambda((W\widehat{Q})_i\beta+l(t_k-))\big)&=&1\nonumber\\
\text{for }\quad k=2,3,\ldots,K.
  \end{eqnarray}

 Thus we have the following  iterative algorithms for computing  $\widehat\beta$ and $\widehat{l}$.
\begin{enumerate}[Step 1.]
\item  Using the observed covariates $Z$ and the instrumental variables $W$, compute $ \widehat{Q}$ using (\ref{qhat}).
\item  Choose  initial estimate $\beta^{(0)}$. Obtain an estimator $\widehat{l}$ for $l$ by solving the equations (\ref{eq5.1}) and (\ref{eq5.2}). %One can find an estimator similar to Breslow (1974) cumulative hazard function estimator
\item  Estimate the regression parameters by solving the equation (\ref{eq3.9}) using $\widehat{l}$ obtained in Step 2.
\item  Set $\widehat\beta$ to be the estimate obtained in previous step and repeat Steps 2-4 until convergence.
\end{enumerate}

Next, we carry out a Monte Carlo simulation study to assess  the finite sample performance of the proposed regression estimators.
%We also find the effect of measurement error variance in regression estimators  by  considering  different values of $\sigma_{v}^{2}$.
We consider the model $l(T)=\log T=-X\beta+e$,  where the hazard function of $e$ is of the form
$\lambda(t)=\frac{\exp(t)}{1+r\exp(t)}$. Here $r=0$ and $r=1$, corresponding to proportional hazards model and proportional odds model.  In simulation, we consider three cases i) with one covariate  and one instrumental variable, ii)  with one covariate  and two instrumental variables and iii) with two covariates and two instrumental variables.

\noindent Case (i):  We generate errors $v$ and $\varepsilon$  from  independent standard normal random variables. The instrumental variable $W$ is generated from exponential distribution with mean 4.  We set $Q=3$ and  $\beta_1=1,2,3$.   The data $X$, $Z$ and $T$ are generated using  model (\ref{3.4}).

\noindent Case (ii):  We generate $v$ and $\varepsilon$   as in Case (i). The instrumental variables $W_1$ and $W_2$ are   generated from exponential distributions with mean 2 and 4, respectively.  We set $Q=(2,3)$ and  $\beta_1=1,2,4$.   The data $X$, $Z$ and $T$ are generated as in Case (i).

\noindent Case (iii):  The components of $v$ and $\varepsilon$ are taken to be independent standard  normal random variables. Instrumental variables $W_1$ and $W_2$ are generated as in Case (ii).  We take $\beta=(2,4), (4,2)$ and $Q=diag(2,5)$.  As in Cases (i) and (ii), we generate the data $X$, $Z$ and $T$  using  model (\ref{3.4}).

In all the cases discussed above, censoring random variable $C$ is simulated from $U(0,c)$ distribution, where $c$  is chosen such that approximately 20\% observations are censored.  The least squares estimator $\widehat{Q}$ of $Q$ is obtained using (\ref{qhat}). Simulation is carried out for  sample sizes $n=10,20,30,40,50$ and are based on ten thousand replications.  Bias and MSE for Case (i)  are report in  Tables 1 and 2 for $r=0$ and $r=1$, respectively. In Tables 3 and 4, we report the results for Case (ii). In Tables 5 and 6, we present  bias and MSE for Case (iii). %We also find the effect of measurement error variance in regression estimators using different values of $\sigma_{v}^{2}$. We illustrate it for Case (i)  and the result is presented in Table 7.

\begin{table}[h]\vspace{-.2cm}
\centering
\caption{Proportional hazard model: \\Bias \& MSE of $\widehat \beta$, $p=q=1$}
\label{t11}
\begin{tabular}{lllllllllllll}
\hline
\multicolumn{1}{c}{} & \multicolumn{2}{c}{$\beta_1=1$} & \multicolumn{2}{c}{$\beta_1=2$} & \multicolumn{2}{c}{$\beta_1=3$}\\ \hline
$n$                  & Bias & MSE           & Bias          & MSE     & Bias          & MSE              \\ \hline
$10$  &0.0065&0.0078&0.0083 &0.0108&0.0562 &0.0451          \\

$20$   & 0.0023& 0.0005& 0.0030& 0.0043&0.0097& 0.0380 \\

$30$   &0.0007& 0.0004&0.0011& 0.0004&0.0022& 0.0005 \\

$40$   &0.0002& 0.0003&0.0005 &0.0003&0.0006& 0.0004 \\

$50$   &0.0001& 0.0003&0.0004&0.0003&0.0006& 0.0003 \\ \hline
\end{tabular}
\end{table}
\begin{table}[htp]\vspace{-.2cm}
\centering
\caption{Proportional odds model: \\Bias \& MSE of $\widehat \beta$, $p= q=1$}
\label{t12}
\begin{tabular}{lllllllllllll}
\hline
\multicolumn{1}{c}{} & \multicolumn{2}{c}{$\beta_1=1$} & \multicolumn{2}{c}{$\beta_1=2$} & \multicolumn{2}{c}{$\beta_1=3$}\\ \hline
$n$                  & Bias & MSE           & Bias          & MSE     & Bias          & MSE              \\ \hline
$10$  &0.0139&0.0030&0.0056&0.0040&0.0120&0.0030          \\

$20$   &0.0065&0.0025&0.0048&0.0038&0.0095&0.0030 \\

$30$   &0.0040&0.0024&0.0038&0.0036&0.0081&0.0032\\

$40$   &0.0022&0.0029&0.0028&0.0034&0.0077&0.0024 \\

$50$  &0.0019&0.0026 &0.0013&0.0020&0.0025&0.0020 \\ \hline
\end{tabular}
\end{table}

\begin{table}[h]\vspace{-.2cm}
\centering
\caption{Proportional hazard model: \\Bias \& MSE of $\widehat \beta$, $p=1,\,  q=2$}
\label{t11}
\begin{tabular}{lllllllllllll}
\hline
\multicolumn{1}{c}{} & \multicolumn{2}{c}{$\beta_1=1$} & \multicolumn{2}{c}{$\beta_1=2$} & \multicolumn{2}{c}{$\beta_1=3$}\\ \hline
$n$                  & Bias & MSE           & Bias          & MSE     & Bias          & MSE              \\ \hline
$10$  &0.0192&0.0006&0.0205 &0.0007&0.0213 &0.0007          \\

$20$   & 0.0177& 0.0005& 0.0188& 0.0006&0.0181& 0.0005 \\

$30$   &0.0178& 0.0005&0.0180& 0.0005&0.0177& 0.0005 \\

$40$   &0.0172& 0.0004&0.0176 &0.0005&0.0172& 0.0005 \\

$50$   &0.0169& 0.0004&0.0168&0.0004& 0.0169& 0.0004 \\ \hline
\end{tabular}
\end{table}
\begin{table}[htp]\vspace{-.2cm}
\centering
\caption{Proportional odds model: \\Bias \& MSE of $\widehat \beta$, $p=1,\,  q=2$}
\label{t12}
\begin{tabular}{lllllllllllll}
\hline
\multicolumn{1}{c}{} & \multicolumn{2}{c}{$\beta_1=1$} & \multicolumn{2}{c}{$\beta_1=2$} & \multicolumn{2}{c}{$\beta_1=3$}\\ \hline
$n$                  & Bias & MSE           & Bias          & MSE     & Bias          & MSE              \\ \hline
$10$  &0.0204&0.0007&0.0205&0.0007&0.0206& 0.0007         \\

$20$   &0.0188&0.0006&0.0186&0.0006&0.0187&0.0006 \\

$30$   &0.0174&0.0005&0.0182&0.0005&0.0182&0.0005 \\

$40$   &0.0171&0.0004&0.0172&0.0005&0.0173& 0.0005\\

$50$   &0.0167&0.0004&0.0171&0.0004&0.0170&0.0004 \\ \hline
\end{tabular}
\end{table}

\begin{table}[h]\label{ph-bias and mse}
\caption{ Proportional hazard model:\\ Bias \& MSE of $\widehat \beta$, $p=q=2$}
\label{t:one}
\begin{small}
\centering
\begin{tabular}{crrrrrrrrr}

    \hline $ n$
    &Bias&MSE&Bias&MSE&Bias&MSE&Bias&MSE\\
        \hline
        &\multicolumn{2}{c}{$\beta_1=2$}&\multicolumn{2}{c}{$\beta_2=4$}&
\multicolumn{2}{c}{$\beta_1=4$}&\multicolumn{2}{c}{$\beta_2=2$}\\
    \hline
     10 & 0.0906&     0.0121&    0.0381    &0.0064&0.0622& 0.0091 &0.0038 &0.0024
     \\
    20 &0.0899& 0.0112 & 0.0359 &0.0021&0.0578& 0.0078& 0.0029 &0.0019
 \\
     30 & 0.0872& 0.0107& 0.0343& 0.0023&0.0533& 0.0063 &0.0027 &0.0016
\\
   40 & 0.0806& 0.0099 &0.0313 &0.0017&0.0469& 0.0061& 0.0024 &0.0016
\\
     50 & 0.0721& 0.0095&0.0270 &0.0021&0.0443 &0.0047& 0.0022 &0.0015
\\
\hline
\end{tabular}
\end{small}
%\label{31}%
\end{table}%

\begin{table}[h]\label{ph-bias and mse}
\caption{ Proportional odds model:\\ Bias \& MSE of $\widehat \beta$, $p=q=2$}
\label{t:one}
\begin{small}
\centering
\begin{tabular}{crrrrrrrrr}

    \hline $ n$
    &Bias&MSE&Bias&MSE&Bias&MSE&Bias&MSE\\
        \hline
        &\multicolumn{2}{c}{$\beta_1=2$}&\multicolumn{2}{c}{$\beta_2=4$}&
\multicolumn{2}{c}{$\beta_1=4$}&\multicolumn{2}{c}{$\beta_2=2$}\\
    \hline
     10 & 0.1046& 0.0136 &0.0335 &0.0017& 0.0660 &0.0082  &0.0073 &0.0024    \\
     20 &0.0996& 0.0122& 0.0317& 0.0019& 0.0654& 0.0079& 0.0068 &0.0023 \\
     30 &  0.0821& 0.0096& 0.0282& 0.0063& 0.0559& 0.0082 &0.0054 &0.0029\\
   40 &  0.0800& 0.0117 &0.0281 &0.0020&0.0555& 0.0071 &0.0048 &0.0015\\
     50 & 0.0770& 0.0091 &0.0280 &0.0015& 0.0473& 0.0066 &0.0047 &0.0015\\
\hline
\end{tabular}
\end{small}
%\label{31}%
\end{table}%

We estimate the coverage probability (CP) and average width (AW) of the confidence interval of the regression parameter for different values of the parameter considered in Case (i). The results for proportional hazards and proportional odds models are given in Tables 7 and 8, respectively. From these tables, we observe that, the coverage probability approaches 0.95 and the average width of the interval decreases as $n$ increases.

%.....PH CP and AW
\begin{table}[h]\label{ph-cp and aw}
\caption{ Coverage probability and average width of the confidence interval of $\beta$ for Case (i) under proportional hazards model}
\label{t:one}
\begin{small}
\centering
\begin{tabular}{crrrrrrrrr}

	\hline
		&\multicolumn{2}{c}{$\beta=1$}&\multicolumn{2}{c}{$\beta=2$}&
\multicolumn{2}{c}{$\beta=3$}\\	\hline
 $ n$
	&CP&AW&CP&AW&CP&AW\\
	\hline	
	10 &0.9422 &0.1002&0.9431&0.1049&0.9429&0.1052 \\
	 20 &0.9437 &0.0826&0.9444&0.0864&0.9449&0.0873 \\
	 30 &0.9447&0.0767&0.9430&0.0776&0.9418&0.0791	\\
	 40&0.9479 &0.0702&0.9452&0.0740&0.9449&0.0756 \\
	 50 &0.9476&0.0669&0.9465&0.0706&0.9467&0.0726\\
		\hline
	\end{tabular}
\end{small}
%\label{31}%
\end{table}%

\begin{table}[h]\label{ph-cp and aw}
\caption{ Coverage probability and average width of the confidence interval of $\widehat \beta$ for Case (i) under proportional odds model}
\label{t:one}
\begin{small}
\centering
\begin{tabular}{crrrrrrrrr}
	\hline
		&\multicolumn{2}{c}{$\beta=1$}&\multicolumn{2}{c}{$\beta=2$}&
\multicolumn{2}{c}{$\beta=3$}\\	\hline
 $ n$
	&CP&AW&CP&AW&CP&AW\\
	\hline	
	10 & 0.9529&	0.1128&	0.9539&	0.1110&0.9529&	0.1106 \\
	 20 & 0.9526&	0.0918&	0.9518&	0.0964 &0.9528&	0.0986 \\
	 30 & 0.9497&	0.0843&	0.9487&	0.0882& 0.9499&	0.0857	\\
	 40&  0.9509&	0.0730&	0.9509&	0.0796 &0.9500	&0.0763 \\
	 50 & 0.9506&	0.0612&	0.9502&	0.0685 &0.9487&	0.0655 	\\
		\hline
	\end{tabular}
\end{small}
%\label{31}%
\end{table}%

%We compare theoretical asymptotic variance and Monte Carlo variance of the estimator of $\beta$  under PH model. In PH model, we obtain the variance-covariance matrix   (\ref{var}) and (\ref{limvar}) as $$\Sigma_{0k}=\Sigma_{1k}=Var\left(\int_{0}^{\infty}(Z-\mu_k(t))dM_k(t)\right), \, k=1,2,$$where $\mu_k(t)=E(Z|\tilde T,\delta=1)$. Hence $\Sigma_k=\Sigma_{0k}^{-1}$. For more details see Chen et al. (2002).  Comparison results are given in Table 9.
%We observe that the estimators of theoretical variance of the estimators and Monte Carlo variance of the estimators  agree each other.

\section{Data Analysis}
The proposed method is illustrated by applying it to an AIDS clinical Trial group (ACTG 175) study reported in Hammer et al. (1996). ACTG 175 is a double-blind randomized clinical trial intended to compare the treatment effects on HIV-infected subjects on the basis of time to progression to AIDS or death. In this trial, 2467 patients with HIV infection were randomly assigned to four treatment groups viz. zidovudine (ZDV) monotherapy, ZDV and didanosine (DDI), ZDV and zalcitabine (ZAL), and didanosine(DDI) monotherapy. Hammer et al. (1996) concluded that the later three treatments slowed down the progression of HIV disease and are superior to treatment with ZDV alone. Hence, we consider two treatment arms: ZDV alone and either of three other treatments.  This data set has been analyzed  by Song and Wang (2014) to shows the applicability of the nonparametric correction approach for estimating the regression parameters in the proportional hazards model when instrumental variables are observed in a subset of the sample. We apply our methodology to a subset of ACTG data $(n= 1725)$ for which instrumental variable is available.

The variables chosen for the analysis are:  Number of days until the first occurrence of (i) a decline in CD4 T cell count of at least 50\% (ii) an event indicating progression to AIDS, or (iii) death, which is the outcome variable of interest.  Covariate measured with error is $\log(CD4)$ count which is measured one week before randomization. It is taken as the baseline CD4 measurement, by assuming the fact that CD4 counts are relatively stable within a short period of time. Instrumental variable chosen for our study is the  $\log(CD4)$ count measured 1-3 weeks prior to randomization.

There are 437  patients in ZDV monotherapy and 1288 patients in the other group. The median follow up time is 33 months and  427 events are observed. It is well known that observed CD4 counts are affected by both instrumental measurement error and biological diurnal fluctuation. Logarithmic transformation is applied to all considered  CD4 counts to achieve approximate constant variance. Figure 1 indicates a moderate positive correlation ($r = 0.66$) between covariate measured with error and instrumental variable.  The proposed method is applied to the data and compared with the naive method. The results are given in Table 9, where both baseline CD4 and treatment are significant and we conclude that the estimates of the proposed method show true treatment effects than the naive estimates. The naive approach is biased due to overestimation based on the  evidences from literature explored this data. The proposed method provides reliable estimates since it takes the measurement error into account. In both methods coefficients of proportional odds model is overestimated and proportional hazards model is a good fit for this data.

\begin{figure}[h]
\centerline{\includegraphics[width=60mm]{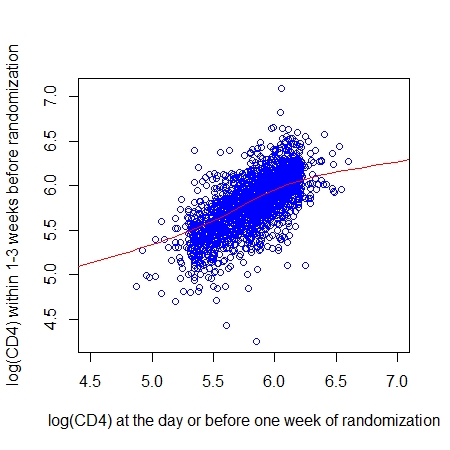}}
\caption{ Scatter plot of log(CD4) within one to three weeks versus up to one week of randomization.}
\label{fig1}
\end{figure}

\begin{table}[]\vspace{-.2cm}
\centering
\caption{Results for ACTG 175 data analysis}
\label{t22}
\begin{tabular}{lllllllllllll}
\hline
\multicolumn{1}{c}{} & \multicolumn{3}{c}{Proportional hazards model}\\
\multicolumn{1}{c}{} & \multicolumn{2}{c}{Treatment} & \multicolumn{2}{c}{log (CD4)} \\ \hline
$Method$                  & Estimate & SE            & Estimate & SE     \\ \hline
Neive &-0.7018&0.1015&-1.2441 &0.1385         \\
Proposed  & -0.6811& 0.1013& -1.5561& 0.1742\\ \hline
\multicolumn{1}{c}{} & \multicolumn{3}{c}{Proportional odds model}\\
\multicolumn{1}{c}{} & \multicolumn{2}{c}{Treatment} & \multicolumn{2}{c}{log (CD4)} \\ \hline
$Method$        & Estimate & SE            & Estimate & SE     \\ \hline
Neive &-0.8314&0.1226&-1.4856& 0.1705         \\
Proposed  &-0.8106&0.1227&-1.8375&0.2066 \\ \hline
\end{tabular}
\end{table}

\section{Conclusion}
In this study, we have proposed an estimation procedure  for handling mis-measured covariates for the linear transformation model in the presence of  instrumental variables.  This approach resolves the measurement error issue in covariates  in many
models including the proportional hazards model and proportional odds model.
 The estimation is done in two steps. In the first step, we estimated mis-measured covariates using instrumental variables.  In the second step, we used these estimates to obtain the estimators of  the regression parameters. The proposed estimators are consistent and have asymptotic normality. Even though the asymptotic variance has complex expression, the consistent estimator can be obtained easily using plug-in method.  Monte Carlo simulation study showed that the proposed method has good finite sample performance in terms of bias  and MSE. Finally, we  illustrated our methodology using an AIDS clinical trial (ACTG 175) data.

As mentioned, Zhou et al. (2022) developed a R package `TransModel' for linear transformation model with censored data. The proposed method can be implemented using the same package as instrumental variable approach allows to estimates the mismeasured covariates. In our study, we  assume that the survival time is conditionally independent of the censoring time given the covariates. Recently, Deresa and Van Keilegom (2021) developed methods for analysing semiparametric transformation model when the  survival time is conditionally dependent of the censoring time, given a set of covariates. They assumed that covariates are measured accurately.  One can develop an  instrumental variable approach to incorporates mismeasured covariates  in their study. Different types of censoring schemes such as current status censoring, double censoring and interval censoring are common in survival studies. The proposed method can be easily  modified to these set up  by suitably constructing martingale based estimating equations.
\section*{Acknowledgements}
Research of Sudheesh is supported by the Mathematical Research Impact Centric Support, 2018-2021, Science and Engineering Research Board, DST, Govt. of India.

\vspace{-0.6in}
\noindent{\bf Proof of Lemma 3.1:} Using the representation  of $\Lambda^*(\widehat{l}(t))$ given in A2 of Chen et al. (2002), we have
  \begin{equation*}
    \Lambda^*(\widehat{l}(t))-\Lambda^*({l}_o(t))=\frac{1}{n}\sum_{i=1}^{n}\int_{0}^{t}
    \frac{\lambda^*(l(s))}{B_2(s)}dM_i(s)+o_p(1/\sqrt n).
  \end{equation*}By Taylor series approximation

  \begin{equation*}
    (\widehat{l}(t)-{l}_0(t))\lambda^*(l_0(t))=\frac{1}{n}\sum_{i=1}^{n}\int_{0}^{t}
    \frac{\lambda^*(l_0(s))}{B_2(s)}dM_i(s)+o_p(1/\sqrt n).
  \end{equation*}
 Differentiating with respect to $Q$,  from the law of large numbers and then using the martingale representation we obtain
 \begin{small}
\begin{equation*}
\frac{\partial}{\partial Q}\widehat{l}(t)\lambda^{*}(l_0(\tilde{T}_i))=-
  \int_{0}^{t}\frac{\lambda_{0}^{*}(l_0(s))}
 {B_2(s)}E\big[W\beta{\lambda}'(WQ\beta+l_0(s))Y(s)\big]dl_0(s)+o_p(1).
  \end{equation*}
  \end{small}Hence we have the first part of the lemma.
 On similar lines, we can prove the second  result stated in the lemma.\\

 \noindent {\bf Proof of Lemma 3.2:}
Consider
\begin{small}
\begin{eqnarray*} &&\hskip-.4in
 \frac{\partial}{\partial \beta}U_{1}(\beta,\widehat l,Q)_{\beta=\beta_a}\\&=&-\sum_{i=1}^{n}
 (WQ)_i\lambda\left((WQ)_i\beta_0+\widehat{l}(\tilde T)\right)\{(WQ)_i+\frac{\partial}{\partial \beta}\widehat{l}(\tilde{T})\}\\&=&-\sum_{i=1}^{n}
 (WQ)_i\lambda\left((WQ)_i\beta_0+\widehat{l}(\tilde T)\right)\\&&\times\Big\{(WQ)_i-
 \int_{0}^{\tilde{T}_i}\frac{\lambda_{0}^{*}\left(l_0(s)E\big[WQ{\lambda}'(WQ\beta_0+l_0(s))Y(s)\big]\right)}
 {\lambda^{*}(l_0(\tilde{T}_i))B_2(s)}dl_0(s)\Big\}+o_p(n).
 \end{eqnarray*}
 \end{small}
Accordingly, in view of the second part of Lemma 3.1, from law of large numbers
\begin{eqnarray*}
 \frac{1}{n}\frac{\partial}{\partial \beta}U_{1}(\beta,\widehat l,Q)_{\beta=\beta_0}\stackrel{p}\rightarrow -\Sigma_{\beta_{0}},
\end{eqnarray*}where $$\Sigma_{\beta_{0}}=\int_{0}^{\tau}E\{(WQ-\mu(t))^T(WQ){\lambda}'\{(WQ)\beta_0+l_0(t)\}Y(t)\}dl_0(t).$$

%In view of the martingale representation, by  law of large numbers,
\noindent {\bf Proof of Lemma 3.3:}
In view of the first  part of Lemma 3.1, by  law of large numbers,
\begin{small}
\begin{eqnarray*}
 &&\hskip-0.6in
\frac{1}{n}\frac{\partial}{\partial Q}U_{1}(\beta,\widehat l,Q)_{Q=Q_0}\\&=&\frac{1}{n}\sum_{i=1}^{n}\int_{0}^{\tau}W_idN_i(t)-\frac{1}{n}
\sum_{i=1}^{n}\int_{0}^{\tau}W_iY_i(t)d\Lambda((WQ_0)_i\beta+\widehat{l}(t))\\&&-\frac{1}{n}
\sum_{i=1}^{n}(WQ_0)_i\lambda((WQ_0)_i\beta+\widehat{l}(t))\big(W_i+\frac{\partial}{\partial Q}\widehat{l}(t)_{Q=Q_0}\big)
%\\&=&-\frac{1}{n}
%\sum_{i=1}^{n}(WQ)_i\lambda((WQ)_i\beta+\widehat{l}(t))\big(W_i+\frac{\partial}{\partial Q}\widehat{l}(t)\big)+o_p(1)
\\&=&- \frac{1}{n}
\sum_{i=1}^{n}(WQ_0)_i\lambda((WQ_0)_i\beta+\widehat{l}(t))\\&&\times \big(W_i-\int_{0}^{\tilde{T}_i}\frac{\lambda_{0}^{*}\left(l_0(s)E\big[W\beta{\lambda}'(WQ_0\beta+l_0(s))Y(s)\big]\right)}
 {\lambda^{*}(l_0(\tilde{T}_i))B_2(s)}dl_0(s)\big)+o_p(1)\\&=&-\Sigma_{Q_{0}}+o_p(1),
\end{eqnarray*}
\end{small}

\noindent{\bf Proof Theorem 3.1:}
Using Taylors theorem, from  (\ref{eq3.9}) we have
 \begin{eqnarray*}
 0&=&U_{1}(\widehat\beta,\widehat l,\widehat{Q})\\
 &=& U_{1}(\beta,\widehat l,\widehat{Q})+(\widehat{\beta}-\beta)\frac{\partial}{\partial \beta}U_{1}(\beta,\widehat l,Q)_{\beta=\beta_a}\\
 &=&A_1+A_2+(\widehat{\beta}-\beta)\frac{\partial}{\partial \beta}U_{1}(\beta,\widehat l,Q)_{\beta=\beta_a},
 \end{eqnarray*} where  $$ A_1=U_{1}(\beta,\widehat l,\widehat{Q})-U_{1}(\beta,\widehat l,Q)\,\,\text{and} \,\,A_2=U_{1}(\beta,\widehat l,Q).$$
 %$\widehat{Q}$ as in (\ref{qhat}).
 Therefore
 \begin{eqnarray*}
 \sqrt{n}(\widehat{\beta}-\beta)\frac{-1}{n}\frac{\partial}{\partial \beta}U_{1}(\beta,\widehat l,Q)_{\beta=\beta_a}&=&\frac{1}{\sqrt{n}}A_1+\frac{1}{\sqrt{n}}A_2.
 \end{eqnarray*}In view of Lemma (\ref{lemma2}), we have the proof of the theorem once we establish the asymptotic normality of   $\frac{1}{\sqrt{n}}A_1$ and $\frac{1}{\sqrt{n}}A_2$.
 %and the converges in probability of the term
% $\frac{-1}{n}\frac{\partial}{\partial \beta}U_{1}(\beta,\widehat l,Q)_{\beta=\beta_a}. $

Consider the term $A_1.$
Using Taylors theorem,
\begin{eqnarray*}
U_{1}(\beta,\widehat l,\widehat{Q})-U_{1}(\beta,\widehat l,Q)=(\widehat{Q}-Q)\frac{\partial}{\partial Q}U_{1}(\beta,\widehat l,Q)_{Q=Q_a}.
\end{eqnarray*}
%\begin{eqnarray*}
%\sqrt{n}(\widehat{Q}-Q)= distribution
%\end{eqnarray*}

By Lemma (\ref{lemma4}), we have $nE(||\widehat{Q}-Q||^2)=O_p(1)$.   Since  $W^TW/n$ converges in probability to a positive definite matrix, by CLT $\sqrt{n}(\widehat{Q}-Q)$ converges in distribution to normal with mean zero and variance-covariance matrix $(W^TW)^{-1}\sigma_{\eta}^{2}$, where $\sigma_{\eta}^2=\sigma_{\varepsilon}^2+\sigma_{v}^2$. In view of Lemma (\ref{lemma3}), $\frac{1}{\sqrt{n}}A_1$ has normal distribution with zero mean vector   and variance-covariance matrix $\Sigma_{1}=\Sigma_{Q_0}(W^TW)^{-1}\Sigma_{Q_0}^T\sigma_{\eta}^{2}$.
The asymptotic behaviour of $A_2$ can be established using martingale theory as analogs to Chen et al. (2002) and
$\frac{1}{\sqrt{n}}A_2\rightarrow N(0,\Sigma_2).$

Next we find $\frac{1}{n}Cov(A_1,A_{2})$. Consider
\begin{small}
\begin{eqnarray}\label{aa}
 \nonumber % Remove numbering (before each equation)
  \frac{1}{n}A_1A_{2}^T&=&\frac{1}{n} U_{1}(\beta,\widehat l,Q)(U_{1}(\beta,\widehat l,\widehat{Q})-U_{1}(\beta,\widehat l,Q))^{T}\\
  &=&\frac{1}{n} U_{1}(\beta,\widehat l,Q)(U_{1}(\beta,\widehat l,\widehat{Q}))^{T}-\frac{1}{n} U_{1}(\beta,\widehat l,Q)(U_{1}(\beta,\widehat l,Q))^{T}.
\end{eqnarray}
\end{small}
Now, the expectation of the second term in the  equation (\ref{aa}) converges to $\Sigma_2$.
Consider
\begin{eqnarray*}
\frac{1}{\sqrt{n}}U_{1}(\beta,\widehat l,\widehat Q)&=&\frac{1}{\sqrt{n}}\sum_{i=1}^{n}\int_{0}^{\tau}(W\widehat{Q})_i[dN_i(t)-
Y_i(t)d\Lambda{(W\widehat{Q})_i\beta_0+\widehat{l}_0(t)}]\\
&=&\frac{1}{\sqrt{n}}\Big\{\sum_{i=1}^{n}\int_{0}^{\tau}(W\widehat{Q})_idM_i(t)\\&&-\sum_{i=1}^{n}(W\widehat{Q})_i
[\Lambda{(W\widehat{Q})_i\beta_0+\widehat{l}_0(t)}-\Lambda{(W\widehat{Q})_i\beta_0+{l}_0(t)}]\Big\}.
\end{eqnarray*}
 Using the representation  of $\Lambda^*(\widehat{l}(t))$ given in A2 of Chen et al. (2002) and using the martingale approximation,  we have
\begin{eqnarray*}
% \nonumber % Remove numbering (before each equation)
 && \sum_{i=1}^{n}\int_{0}^{\tau}(W\widehat{Q})_idM_i(t)-\sum_{i=1}^{n}(W\widehat{Q})_i
[\Lambda{(W\widehat{Q})_i\beta_0+\widehat{l}_0(t)}-\Lambda{(W\widehat{Q})_i\beta_0+{l}_0(t)}]\\
&&\qquad=\sum_{i=1}^{n}\int_{0}^{\tau}((W(W^TW)^{-1}W^TZ)_i-\mu_{\widehat{Q}}(t))dM_i(t)+o_p(\sqrt{n}),
\end{eqnarray*}which gives
\begin{eqnarray*}
\frac{1}{\sqrt{n}}U_{1}(\beta,\widehat l,\widehat Q)=\frac{1}{\sqrt{n}}\sum_{i=1}^{n}\int_{0}^{\tau}((W(W^TW)^{-1}W^TZ)_i-\mu_{\widehat{Q}}(t))dM_i(t)+o_p(1),
\end{eqnarray*}
Using the same arguments as above, we obtain
\begin{eqnarray*}
% \nonumber % Remove numbering (before each equation)
 \frac{1}{\sqrt{n}}U_{1}(\beta,\widehat l, Q)&=& \frac{1}{\sqrt{n}}
\sum_{i=1}^{n}\int_{0}^{\tau}((W{Q})_i-\mu_{Q}(t))dM_i(t)+o_p(1).
\end{eqnarray*}
Hence the expectation of the first term in (\ref{aa}) converges to $\Sigma_{12}$. This complete the proof.

\end{document}